\documentclass[iop]{emulateapj}
\journalinfo{Accepted for publication in ApJ Letters}
\submitted{Accepted for publication in ApJ Letters, \today}
\usepackage{graphicx}

\usepackage{amsmath}
\usepackage[]{natbib}
\usepackage[usenames]{color} 
\usepackage{ulem}
\def\apj{{Astroph. J.}}
\def\apjs{{Astroph. J. Suppl.}}
\def\aj{{Astron. J.}}
\def\mnras{{MNRAS}}

\shorttitle{The Shape of the Milky Way Dark Matter Halo}

\shortauthors{Loebman et al.~}
\begin{document}

\title{Constraints on the Shape of the Milky Way Dark Matter Halo
                    from Jeans Equations Applied to SDSS Data}

\author{Sarah R. Loebman\altaffilmark{1}}
\author{\v{Z}eljko Ivezi\'{c}\altaffilmark{1}}
\author{Thomas R. Quinn\altaffilmark{1}}
\author{Fabio Governato\altaffilmark{1}}
\author{Alyson M. Brooks\altaffilmark{2}}
\author{Charlotte R. Christensen\altaffilmark{3}}
\author{Mario Juri\'{c}\altaffilmark{4}}

\altaffiltext{1}{Astronomy  Department, University of  Washington, 
                 Box 351580, Seattle, WA 98195-1580;       
		 {\tt sloebman@astro.washington.edu}}
\altaffiltext{2}{Grainger Postdoctoral Fellow, Department of Astronomy, 
                 University of Wisconsin}
\altaffiltext{3}{Theory Fellow, Astronomy Department, 
                University of Arizona, Tucson}
\altaffiltext{4}{LSST Corporation, 933 North Cherry Avenue, 
                 Tucson, AZ 85721}

\begin{abstract} 
We search  for evidence of dark  matter in the Milky  Way by utilizing
the stellar number density distribution and kinematics measured by the
Sloan Digital  Sky Survey  (SDSS) to heliocentric  distances exceeding
$\sim$10  kpc.  We employ  the cylindrically  symmetric form  of Jeans
equations and  focus on the  morphology of the  resulting acceleration
maps,  rather than  the normalization  of the  total mass  as  done in
previous, mostly local, studies.  Jeans equations are first applied to
a  mock catalog  based  on a  cosmologically derived  $N$-body~$+$~SPH
simulation,  and  the known  acceleration  (gradient of  gravitational
potential)  is successfully  recovered.  The  same simulation  is also
used to quantify the impact  of dark matter on the total acceleration.
We  use Galfast,  a  code designed  to  quantitatively reproduce  SDSS
measurements and  selection effects,  to generate a  synthetic stellar
catalog.   We  apply  Jeans  equations  to this  catalog  and  produce
two-dimensional maps of stellar  acceleration.  These maps reveal that
in a  Newtonian framework, the implied  gravitational potential cannot
be explained by visible matter alone.  The acceleration experienced by
stars  at galactocentric  distances  of $\sim$20  kpc  is three  times
larger  than what  can be  explained  by purely  visible matter.   The
application of an analytic method  for estimating the dark matter halo
axis ratio to SDSS data implies an oblate halo with $q_{DM} = 0.47 \pm
0.14$ within the same distance range.  These techniques can be used to
map the  dark matter halo to  much larger distances  from the Galactic
center using upcoming deep optical surveys, such as LSST.
\end{abstract}

\keywords{stars:  kinematics and  dynamics ---  stars:  statistics ---
Galaxy:  general  ---  Galaxy:  kinematics and  dynamics  ---  Galaxy:
structure --- Galaxy: halo}


\section{Introduction}
\label{s:intro.tex}

Determining the  dark matter  content of the  Milky Way  has important
implications for fields ranging  from theories of galaxy formation and
evolution  to experimental  physics.  Lately,  there has  been renewed
interest in  an old concept --  applying Jeans equations  to stars in
the   Milky   Way   to   infer  the   underlying   mass   distribution
\citep{Jeans1915,  Oort1932}.  This technique  statistically estimates
the  gravitational  potential  using  observable  stellar  kinematics,
rather than accelerations that are rarely detectable.

Jeans equations follow from  the collisionless Boltzmann equation; for
a  detailed  derivation  see  \citet{Binney1987}.   Using  cylindrical
coordinates and assuming an axi-symmetric and steady-state system, the
accelerations in the radial ($R$) and vertical ($Z$) directions can be
expressed  in  terms  of  observable quantities:  the  stellar  number
density distribution, $\nu$,  the mean azimuthal (rotational) velocity
$\overline{v_{\phi}}$,     and      four     velocity     dispersions,
$\sigma_{\phi\phi}$,  $\sigma_{RR}$, $\sigma_{ZZ}$,  and $\sigma_{RZ}$
(all as functions of $R$ and $Z$), as
\begin{eqnarray}
\label{eq1}
a_{R} =& \sigma_{RR}^2 \times \frac{\partial (\ln\nu)}{\partial R} + 
         \frac{\partial \sigma_{RR}^2}{\partial R} + 
         \sigma_{RZ}^2 \times \frac{\partial (\ln\nu)}{\partial Z} + \\
       & \frac{\partial \sigma_{RZ}^2}{\partial Z} + 
         \frac{\sigma_{RR}^2 }{R} - 
         \frac{\sigma_{\phi\phi}^2 }{R} - 
         \frac{{ \overline{v_\phi} }^2 }{R},  \nonumber
\end{eqnarray}
\begin{eqnarray}
\label{eq2}
a_{Z} =& \sigma_{RZ}^2 \times \frac{\partial (\ln\nu)}{\partial R} +
         \frac{\partial \sigma_{RZ}^2}{\partial R} +
         \sigma_{ZZ}^2 \times \frac{\partial (\ln\nu)}{\partial Z} + \\
       & \frac{\partial \sigma_{ZZ}^2}{\partial Z} +
         \frac{\sigma_{RZ}^2 }{ R} \nonumber. 
\end{eqnarray}
Given  accelerations $a_{R}(R,Z)$  and $a_{Z}(R,Z)$,  {\it  i.e.~} the
gradient of the gravitational  potential, the dark matter contribution
can be  estimated after accounting  for the contribution  from visible
matter.

Traditionally,  such  studies  were  limited  by  data  to  the  solar
neighborhood  \citep[within $\sim$150  pc, {\it e.g.~}][]{Kapteyn1922,
Oort1960,  Bahcall1984}.  The  main conclusion  drawn from  such local
studies is  that dark matter contributes  a small (of  the order 10\%)
but significantly detected fraction  of mass in the solar neighborhood
\citep[corresponding  to about  0.01 $M_\sun$  pc$^{-3}$, or  0.38 GeV
cm$^{-3}$; ][]{Kuijken1989c,Creze1998,Holmberg2000}.

Several groups  have extended these  studies to a few  kiloparsec from
the plane  of the disk  \citep{Kuijken1991, Siebert2003, Holmberg2004,
Smith2012, Bovy2012a}.  Recently, \citet{Garbari2012} used a sample of
2000 K dwarf  stars that extend to  1 kpc above the plane  of the disk
and estimated the local  dark matter density distribution ${\rho}_{dm}
= (0.022 \pm  0.015) \, M_\sun \, pc^{-3}$.   Using kinematic data for
$\sim$400 thick disk stars at distances of a few kpc from the Galactic
plane    from   \citet{MoniBidin2012},    \citet{Bovy2012}   estimated
${\rho}_{dm} = (0.008 \pm 0.002) \, M_\sun \, pc^{-3}$.

It has been difficult to extend these measurements to distances beyond
a few kiloparsec  from the solar neighborhood \citep{vanderMarel1991}.
Recently, using  a sample of $\sim$2,500 blue  horizontal branch stars
from SDSS  DR6, \citet{Xue2008} found  an estimate of the  Milky Way's
circular velocity curve at $\sim$60  kpc that implied the existance of
dark matter.   Using a spherical approximation of  Jeans equations and
extending    the    analysis    of   the    \citet{Xue2008}    sample,
\citet{Samurovic2011} also concluded  that the Newtonian model without
dark matter cannot fit  the observed velocity dispersion profile. They
also tested various MOND models  and concluded that these fit the data
as well.

Here we  extend these studies and introduce  a novel multi-dimensional
application of Jeans equations made  possible by the Sloan Digital Sky
Survey\footnote{www.sdss.org} data \citep[][hereafter SDSS]{SDSS2000}.
Due  to  substantial  SDSS   sky  coverage  and  accurate  multi-color
photometry to  faint limits,  the stellar number  density distribution
and stellar  kinematics were mapped  out using numerous  main sequence
stars   detected   to  galactocentric   distances   of  $\sim$20   kpc
\citep{Juric2008,Ivezic2008,Bond2010}.   The extent  of these  maps is
sufficiently  large  that  it   is  possible  to  investigate  stellar
acceleration via Jeans equations.  Most importantly, while the spatial
derivatives  of the  velocity  dispersion are  extremely difficult  to
constrain  from the  local solar  neighborhood, they  can  be directly
measured  using  SDSS  data.   We  discuss  here  the  following  main
questions:
\begin{itemize}
\item How does  the inclusion or exclusion of  a dark matter component
affect the morphology of the stellar acceleration maps?
\item Is  it possible to  recover the known accelerations  by applying
          Jeans equations  to a realistic simulated galaxy  that has a
          merger history and is not perfectly axi-symmetric?
\item Are stellar acceleration  maps derived from SDSS data consistent
          with expectations based only on visible matter?
\end{itemize}

This  paper provides  a  brief  summary of  our  analysis; a  detailed
discussion will be presented elsewhere (Loebman et al.~, in prep).  In
\S\ref{s:background}, we describe the simulation employed in this work
and answer the first two questions.   The main result of this work, an
answer to  the third question, is presented  in \S\ref{s:results}.  We
summarize and discuss our results in \S\ref{s:conclusion}.

\section{Testing the Methodology}
\label{s:background}

\subsection{N--body$+$SPH Simulation}
\label{s:simulation}

\begin{figure}[!h]
\epsscale{1}
\hskip -.15 in
  \plotone{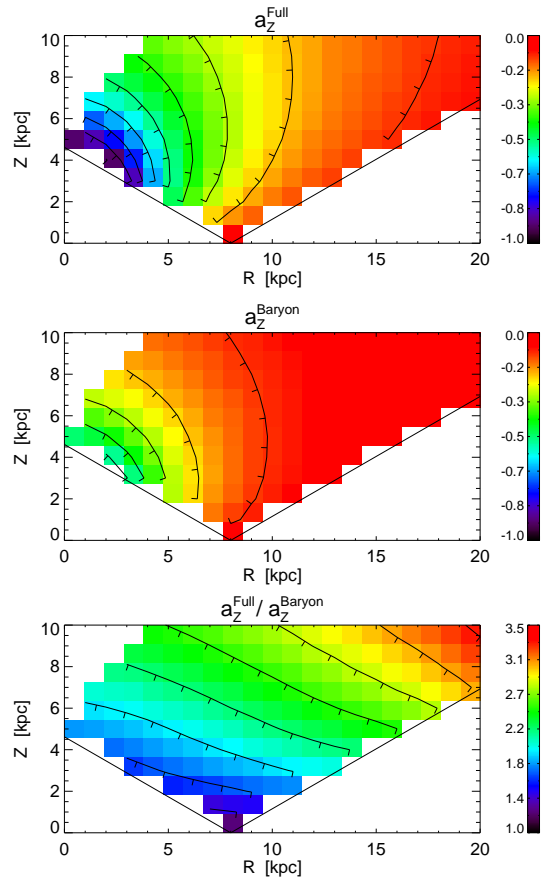}            
  \caption{A comparison of the  acceleration in the $Z$ direction when
  all  contributions   are  included  (star,  gas,   and  dark  matter
  particles;  top panel)  to the  result without  dark  matter (middle
  panel).    The    acceleration   is    expressed    in   units    of
  $2.9\times10^{-13}$ km/s$^2$. The  ratio of the
  two maps is  shown in the bottom panel.  The  importance of the dark
  matter increases with  the distance from the origin;  at the edge of
  the volume probed by SDSS  ($R\sim20$ kpc, $Z\sim10$ kpc), the total
  acceleration in the analyzed simulation is about 3 times larger than
  the contribution  from the visible  matter. The maps are  limited to
  the volume explored by SDSS data, as indicated by the diagonal lines
  encapsulating the colored pixels.}
\label{f:ratio}
\end{figure}

To test the Jeans equation approach,  we apply our analysis tools to a
simulation with  {\it known} accelerations  and velocities.  We  use a
cosmologically  derived \citep[WMAP3,][]{Spergel2003}  Milky Way--mass
galaxy evolved for $13.7$  Gyr using the parallel $N$--body$+$SPH code
GASOLINE  \citep{Wadsley2004}, which  contains realistic  gas, cooling
and  stellar feedback \citep{Stinson2006,  Shen2009, Christensen2012}.
We  track  the  galaxy's  formation  and evolution  using  the  volume
renormalization      technique      \citep{Katz1993,      Pontzen2008,
Governato2012}.  Our  simulated galaxy includes a  stellar halo, which
is built  up primarily  during the merging  process in  a $\Lambda$CDM
cosmology \citep[{\it e.g.~} ][]{Bullock2005,Zolotov2009}.

GASOLINE simultaneously calculates  the potential and the acceleration
that  particles feel;  force  calculations are  consistent with  other
state-of-the-art  cosmological  gas-dynamical codes  \citep{Power2003,
Scannapieco2012}.  The  typical RMS acceleration  error is $\sim$~$0.2\%$
\citep{Wadsley2004}.    The   average   stellar   particle   mass   is
$\sim$~$5800$  $M_\sun$   and  the   dark  matter  particle   mass  is
$1.3\times10^5$ $M_\sun$, with a dark matter softening length of $173$
pc.  At  redshift of zero, the  simulated galaxy has  a virial radius,
defined  at  $\rho/\rho_{crit}=100$, of  $227$  kpc \citep[versus  the
Milky   Way's  $200$   kpc, see][]{Boylan2011}   and  a   virial  mass   of
$7\times10^{11}$    $M_\sun$    \citep[versus    the    Milky    Way's
$1.0^{+0.3}_{-0.2} \times10^{12}$ $M_\sun$, see][]{Xue2008, Klypin2002}, of
this, 7\%  is in gas,  6\% is  in stars, and  87\% is in  dark matter.
Dark  matter  consists  of  36\%  of  the total  mass  in  the  region
corresponding to the solar neighborhood ($7 \le R/kpc \le 9$ \& $0 \le
Z/kpc  \le 1$).   The simulated  galaxy is  approximately rotationally
symmetric (a total matter axis ratio $b:a > 0.9$ within 100 kpc, and a
stellar matter axis ratio $b:a >  0.95$ at $R$=10 kpc), has a $R$-band
disk scale length of $\sim$3.1 kpc \citep[versus the Milky Way's $3.6$
kpc, see][]{Juric2008}  and  corresponding  bulge  to disk  ratio  of  0.33
\citep{Jonsson2006},  and  the circular  velocity  at  2.2 disk  scale
lengths  is  $\sim$208  km/s   \citep[versus  the  Milky  Way's  $220$
km/s, see][]{Xue2008}.    We  draw   direct  comparisons   between  the
simulation and the Milky Way as these structural parameters are within
15\% of one another.

For the sake of brevity, we focus our presentation on the acceleration
in the  $Z$ direction, $a_{Z}$; detailed analysis  of the acceleration
in  the $R$  direction, $a_{R}$,  will  be presented  in a  subsequent
paper.  The top panel of Figure~\ref{f:ratio} shows $a_{Z}$ within the
simulation.  For  comparison, the middle panel shows  an analogous map
when only contributions from star  and gas particles are included.  As
is evident, there are substantial differences in the morphology of the
two maps; the bottom panel demonstrates that the effect of dark matter
on  the resulting  acceleration  increases quickly  towards the  outer
parts  of the  galaxy;  for  example, the  ratio  of accelerations  is
doubled by $R$=8 kpc and $Z$=6 kpc. These distances are probed by SDSS
-- hence our results suggest that the effect of dark matter on stellar
acceleration may be uncovered in SDSS data.


\subsection{Application of Jeans Equations to the Simulation}
\label{s:methods}

Here  we verify  that  the known  gravitational  accelerations in  the
simulation can  be recovered  using Jeans  equations.  We  bin the
star particles into  1 kpc by 1 kpc rectilinear  pixels in $R-Z$ space
and limit our analysis to bins containing at least 100 star particles.
Using  weights proportional  to the  mass  of each  star particle,  we
create  maps  of  $\ln(\nu)$  and kinematic  quantities.   The  spatial
derivatives  of these  quantities at  the position  of each  pixel are
computed by  first fitting  a second order  polynomial in $R$  and $Z$
using  the  eight  neighboring  pixels, and  then  taking  derivatives
analytically;  edge pixels  are discarded  to minimize  the  impact of
fitting errors.

\begin{figure}[!h]
\epsscale{1}
  \plotone{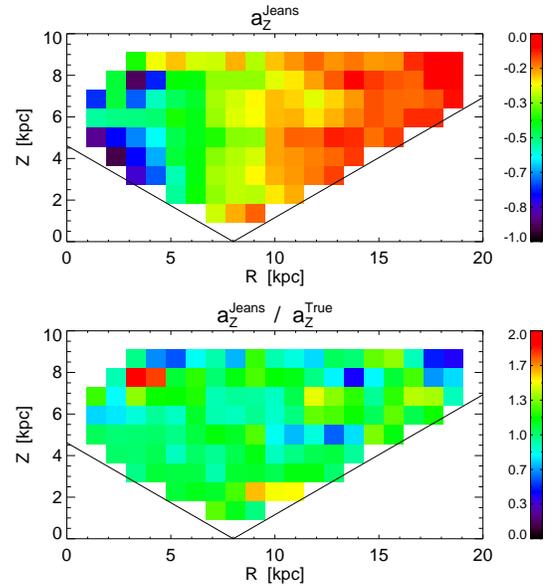}     
  \caption{Top panel:  the acceleration  in the $Z$  direction derived
  using  eq.~\ref{eq2} and expressed  in units  of $2.9\times10^{-13}$
  km/s$^2$.  Bottom panel:  the ratio of  the top
  panel to  the true acceleration in  the $Z$ direction  (top panel of
  Figure~\ref{f:ratio}).}
\label{f:jeans_works}
\end{figure}

We  compare  the acceleration  in  the  $Z$  direction computed  using
eqs.~\ref{eq1}   and   \ref{eq2}   to   the   true   acceleration   in
Figure~\ref{f:jeans_works}.  The  recovered acceleration map  is close
to  the true map:  the distribution  of pixel  values from  the bottom
panel is  centered on  1.05 with a  root-mean-square scatter  (rms) of
0.24.  Similar  agreement  is obtained  for  the  acceleration in  $R$
direction, which is centered on 1.02 with a rms value of 0.38.

We note that Jeans equations recover meaningful acceleration maps even
though the simulation is  neither perfectly rotationally symmetric nor
steady state.  Given this precedence,  we apply the same  technique to
synthetic SDSS data.

\section{Results}
\label{s:results}

\begin{figure}[!h]
\epsscale{1}
  \plotone{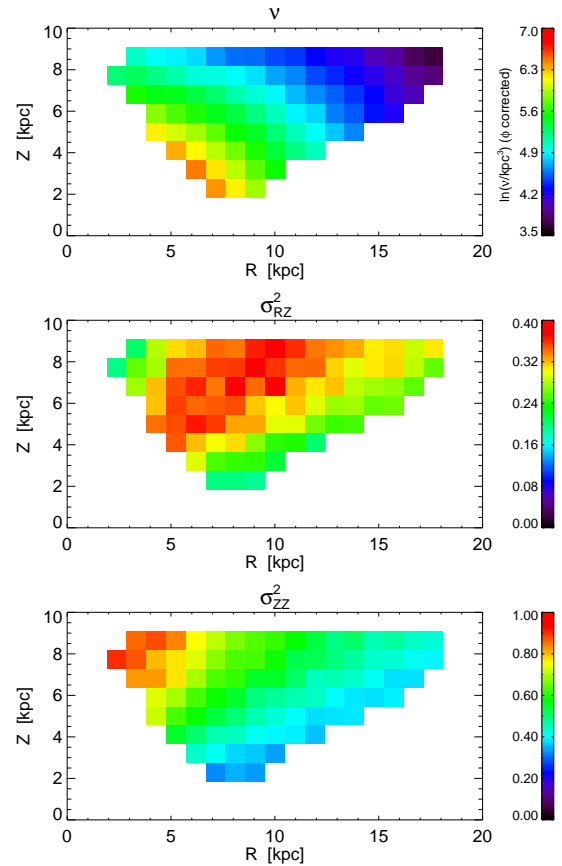}
  \caption{The three  observables as  implied by  SDSS  data (number
density  distribution   in  the  top  panel,  and   the  two  velocity
dispersions in the other panels, as marked, both expressed in units of
$2\times10^4$  (km/s)$^2$)  that are  required  to compute  $a_Z(R,Z)$
using eq.~\ref{eq2}.   The pixels at  $Z<1$ kpc are unreliable  due to
SDSS  saturation  at $r=14$;  additionally  we  restrict  our fits  to
regions with 8 adjacent pixels.}
\label{f:moments}
\end{figure}

\begin{figure}[th!]
\epsscale{1}
  \plotone{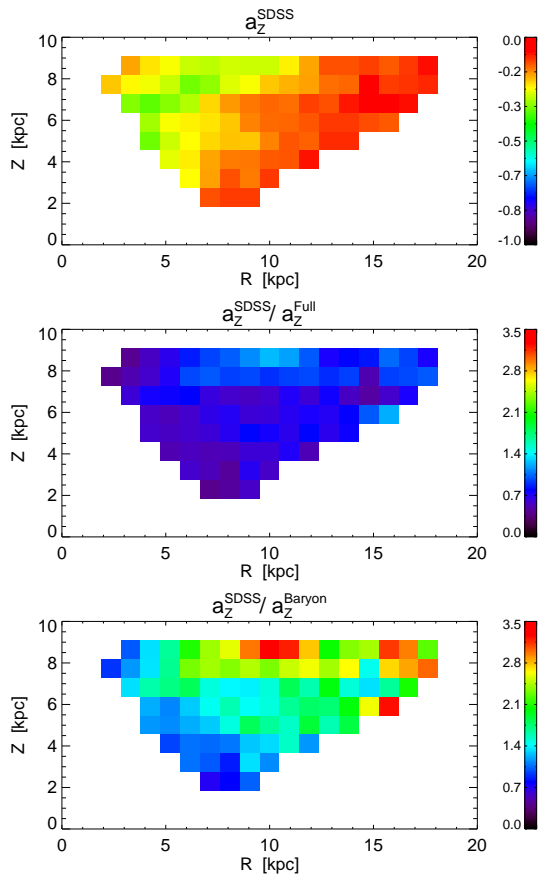}
  \caption{The  results of  applying  Jeans  equations  to the  SDSS
observations simulated  using Galfast.  The  top panel shows a  map of
the  acceleration   in  the  $Z$  direction  expressed   in  units  of
$2.9\times10^{-13}$ km/s$^2$.   The middle and bottom  panels show the
ratio of the map from the top panel and the two model-based maps shown
in the top two panels in Figure~\ref{f:ratio}.}
\label{f:Galfast}
\end{figure}

\begin{figure}[!th]
\epsscale{1}
  \plotone{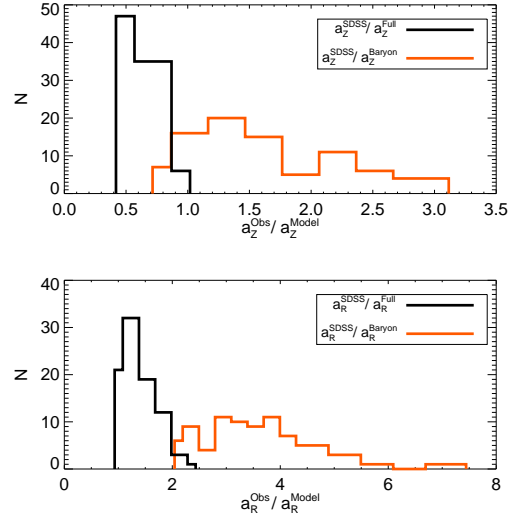}
  \caption{A pixel-by-pixel  comparison of  the  acceleration values
implied by  the SDSS data and  the two model  predictions that include
(black lines) and  do not include (red lines)  contributions from dark
matter.   The top  panel corresponds  to the  $a_Z$  acceleration maps
shown  in Figure~\ref{f:Galfast}  and the  bottom panel  to  the $a_R$
acceleration maps.   The model-based acceleration maps  that include a
dark matter contribution provide a significantly better description of
the acceleration maps derived from the SDSS data.}
\label{f:hist}
\end{figure}

A  direct  application  of  Jeans  equations to  SDSS  data  would  be
complicated because of the complex selection effects and the impact of
substructure.  Instead,  we employ a catalog generated  using the code
Galfast,  which  is  designed  to quantitatively  reproduce  the  SDSS
measurements,  volume  coverage,   selection  and  other  instrumental
effects.  It incorporates smooth models for the stellar number density
distribution,  metallicity   distribution  and  kinematics   based  on
tomographic          analysis           of          SDSS          data
\citep{Juric2008,Ivezic2008,Bond2010}.  The most important ingredients
for  this study are  the best-fit  power-law halo  given by  eq.~24 in
\citet{Juric2008}, with  power-law index $n_H=2.77\pm0.2$,  axis ratio
$q_H=0.64\pm0.1$,  and halo  velocity ellipsoid  that is  invariant in
spherical  coordinates  \citep{Bond2010},  with  $\sigma_r=141$  km/s,
$\sigma_\theta=75$  km/s, and $\sigma_\phi=85$  km/s.  While  the same
analytic models could be  used to directly generate acceleration maps,
we use mock catalogs instead to account for shot noise due to a finite
number of stars and edge effects  and to ensure that the same analysis
code verified on the simulation is used when processing real data.  We
emphasize that  the cylindrical symmetry  built into Galfast  is fully
consistent with  SDSS data \citep[after local  substructure is masked,
see][]{Juric2008}.

Using Galfast, we  generate a flux-limited catalog with $14  < r < 21$
and mimic  the SDSS  sky footprint by  only considering  high Galactic
latitudes ($|b|>30^\circ$).  A halo-like  sample of 0.61 million stars
is selected using $M_r \ge 4$ and  $0.25 < g-r < 0.35$. This sample is
dominated by low-metallicity main  sequence F stars. The catalog lists
true  positions, absolute magnitudes,  velocities and  metallicity, as
well  as  corresponding  simulated  SDSS observations  convolved  with
measurement errors.

Figure~\ref{f:moments} shows  the three observables  constructed using
the  Galfast  catalog  that   are  required  to  compute  $a_Z$  using
eq.~\ref{eq2}.   As expected, both  velocity dispersions  are smoothly
varying,  and the  $\sigma^2_{RZ}$  is  at most  40\%  of the  maximal
$\sigma^2_{ZZ}$ value.

The top panel of Figure~\ref{f:Galfast} shows the resulting $a_Z$ map.
The middle  and bottom panels show  the ratio of this  map compared to
the simulation data in the two maps shown in the top and middle panels
in  Figure~\ref{f:ratio}.  It  is  evident that  the acceleration  map
implied by  SDSS data is  much closer to the  model-based acceleration
map  that includes contributions  from both  baryons and  dark matter.
When dark matter is not  included, the model-based map has a different
shape  than the  map derived  from  SDSS data:  it under-predicts  the
observed acceleration by roughly a factor  of three at $R \sim 17$ kpc
and $Z \sim 9$ kpc. A  pixel-by-pixel comparison of the two ratio maps
is shown  in the the top  panel in Figure~\ref{f:hist};  the fact that
the {\it  shape} of  the observed acceleration  map is  better matched
when dark  matter is  included is seen  as a much  narrower histogram.
These  results  are   not  unique  to  $a_Z$;  the   bottom  panel  in
Figure~\ref{f:hist} shows the same  result for $a_R$, with even larger
deviations between the observed map  and the model-based map that does
not include dark matter contribution.

Our  preliminary  analysis  of  other simulations  indicate  that  the
observed acceleration  map cannot  be reproduced by  simply increasing
the  amount  of  baryons  and  not  including  dark  matter:  the  key
difficulty is to reproduce  the strong acceleration at $10$---$20$ kpc
from the Galactic center while simultaneously maintaining the shape of
the acceleration map throughout the probed volume.

\section{Discussion and Conclusions}
\label{s:conclusion}

We have  shown that the kinematics of  stars can be used  to probe the
dark  matter distribution  in the  Milky  Way.  To  do so,  we used  a
cosmologically derived  $N$--body$+$SPH Milky Way--like  simulation to
demonstrate  that  Jeans  equations  are  capable  of  recovering  the
underlying  gravitational potential despite  small deviations  from an
implied  steady-state axi-symmetric  system. The  same  simulation was
also used  to estimate the  dark matter contribution to  the resulting
acceleration maps.

We  also generated  a synthetic  SDSS  catalog using  Galfast, a  code
designed to  quantitatively reproduce measurements  of spatial density
and kinematics  of Milky Way stars,  as well as  SDSS volume coverage,
selection  and  other  instrumental  effects. With  this  catalog,  we
created  acceleration maps  using Jeans  equations. The  morphology of
these  maps  provides  strong   evidence  for  the  presence  of  dark
matter. This evidence  is not sensitive to the  overall dark matter to
baryon  mass   ratio  (that  is,   the  {\it  normalization}   of  the
acceleration  maps), which can  be considered  as a  fine-tuning model
parameter, but  is robustly derived  from the {\it shape}  of observed
acceleration maps.

While the  method presented here does  not yet provide  errors for the
total estimate  of dark matter  in the Milky  Way, it does  provide an
expanded  view of  the distribution  of dark  matter beyond  the solar
neighborhood.   Significantly,  it  allows  us  to  consider  how  the
distribution of  dark matter evolves  both radially and  vertically in
the context  of a large,  observationally motivated dataset.   Given a
large  number   of  simulations   with  different  dark   matter  halo
properties, it will  be possible to constrain the  overall dark matter
to baryonic  mass ratio and the  shape of the Milky  Way's dark matter
halo.

Before such a suite of  simulations is generated, we can approximately
estimate the shape  of the dark matter halo  using the method proposed
by \cite{vanderMarel1991}.   He showed that for an  oblate dark matter
halo  with a  density distribution  similar to  a  singular isothermal
sphere, it is  possible to derive the dark matter  axis ratio from the
stellar halo axis ratio and the velocity ellipsoid (via application of
Jeans equations;  see his eqs.   1 and 21  and Figure 1).   Using SDSS
measurements  for  the  stellar  halo  axis  ratio  and  the  velocity
ellipsoid  (see \S\ref{s:results}),  we  estimated that  the minor  to
major axis ratio of the Milky Way's dark matter halo is $q_{DM} = 0.47
\pm 0.14$.  Compared to the stellar halo axis ratio, $q_H=0.64\pm0.1$,
the dark matter halo is more oblate.

The  $N$-body$+$SPH   simulation  used  in  this   work  supports  the
properties  of the dark  matter halo  assumed in  the van  der Marel's
method;  in  particular,  its  density  decrease  is  consistent  with
expected $1/[R^2 + (Z/q_{DM})^2]$ within  30 kpc from the origin (with
the power-law  index decreasing  from $-2$ to  about $-2.8$  at larger
distances).  At the  same time,  the stellar  halo is  consistent with
$n_H=2.77$ measured  by SDSS in the  same distance range.  While it is
premature to declare $q_{DM} =  0.47 \pm 0.14$ as a robust measurement
of the dark  matter halo shape, it is  encouraging that the simulation
is  at  least qualitatively  consistent  with  SDSS  data in  so  many
aspects.

Moreover, it is  possible to go beyond Jeans  equations to use stellar
kinematics to  probe the full  phase space distribution of  stars.  As
\citet{Valluri2012}  demonstrated, orbital  spectral  analysis can  be
used to determine not only the shape of the inner halo, but the global
shape  of  the Galactic  halo  as  well.  Their technique  provides  a
complementary tool  to the method presented here  for constraining the
potential and the stellar distribution function.

We note, there  is a considerable range of Milky  Way dark matter halo
axis  ratios presented  in  the literature.  The  reported axis  ratio
ranges from $>$0.7 at the  low end \citep{Ibata2001} to $5$/$3$ at the
high   end   \citep{Helmi2004}.    Sometimes,  triaxial   models   are
incorporated, such as by  \citet{Law2010} who used observations of Sgr
tidal stream and N-body modeling to  conclude that the Milky Way has a
triaxial dark  matter halo  that is nearly  an oblate  ellipsoid whose
minor    axis    is     contained    within    the    Galactic    disk
plane. \citet{Lux2012}  used satellite galaxies to find  the Milky Way
dark   matter  halo   is  oblate,   while   \citet{Banerjee2011}  used
observations of HI  and claimed variation of the  axis ratio, with the
halo increasingly prolate at large distances.

The full  promise of  various methods for  estimating the  geometry of
dark matter halos will be reached by upcoming next-generation surveys,
such as Gaia \citep{PerrymanGaia} and LSST \citep{LSSToverview}.  Gaia
will provide measurements of geometric distances and kinematics with a
similar faint flux limit as  SDSS, but with much smaller errors.  LSST
will  obtain  photometric   distances  and  kinematics  of  comparable
accuracy  to those  of Gaia  at Gaia's  faint limit,  but  extend them
deeper by  about 5~mag.  With LSST  and Gaia, it  will be  possible to
extend Galactic  potential studies such  as the one presented  here to
distances roughly 10 times larger  than those possible with SDSS data,
revolutionizing our understanding of the Milky Way in the process.


\acknowledgements 
\label{s:ack}
SL  and  \v{Z}I  acknowledge  support  by NSF  grants  AST-0707901  \&
AST-1008784 to the University  of Washington, by NSF grant AST-0551161
to  LSST for  design and  development  activity, and  by the  Croatian
National Science  Foundation grant O-1548-2009.   Resources supporting
this work were  provided by the NASA High-End  Computing (HEC) Program
through  the  NASA  Advanced  Supercomputing (NAS)  Division  at  Ames
Research Center.  Galfast computations were performed on Hybrid at the
Physics  Department, University  of  Split, financed  by the  National
Foundation for Science, Higher Education and Technological Development
of the  Republic of Croatia.   FG acknowledges support from  NSF grant
AST-1108885; AB acknowledges support  from the Grainger Foundation. SL
acknowledges support from the Washington NASA Space Grant Consortium.


\end{document}